\documentclass[runningheads]{llncs}

\usepackage{graphicx}
\usepackage{newtxtext,newtxmath,amsmath}
\PassOptionsToPackage{hyphens}{url}
\usepackage{comment}
\usepackage{todonotes}
\usepackage{xcolor}
\definecolor{light-gray}{gray}{0.95}
\usepackage{tcolorbox}
\usepackage{multirow}
\usepackage{graphicx}
\usepackage[all]{nowidow}
\usepackage[colorlinks=true,urlcolor=blue,citecolor=blue,linkcolor=blue]{hyperref}

\DeclareRobustCommand*\circled[1]{\tikz[baseline=(char.base)]{ \node[shape=circle,draw,color=white,fill=black,inner sep=.75pt] (char){#1};}}

\newcommand{\revision}[1]{#1}

\newcommand{\abbas}[1]{#1}

\begin{document}
%
\title{An Analysis of Malware Trends in Enterprise Networks}
\titlerunning{An Analysis of Malware Trends in Enterprise Networks}
%
\author{Abbas Acar\inst{1},
Long Lu \inst{2},
A. Selcuk Uluagac \inst{1},
Engin Kirda \inst{2} 
}

\authorrunning{A. Acar et al.}
%
\institute{Florida International University \\ \email{\{aacar001,suluagac\}@fiu.edu} \and 
Northeastern University \\ \email{l.lu@northeastern.edu,ek@ccs.neu.edu } 
} 

\maketitle              

\begin{abstract} We present an empirical and large-scale analysis of malware
samples captured from two different enterprises from 2017 to early 2018. Particularly, we perform threat vector, social-engineering, vulnerability and
time-series analysis on our dataset. Unlike existing malware studies, our
analysis is specifically focused on the recent enterprise malware samples. First
of all, based on our analysis on the combined datasets of two enterprises, our
results confirm the general consensus that AV-only solutions are not enough for
real-time defenses in enterprise settings because on average 40\% of the malware
samples, when first appeared, are not detected by most AVs on VirusTotal or not
uploaded to VT at all (i.e., never seen in the wild yet). Moreover, our analysis
also \abbas{shows} that enterprise users transfer documents more than \revision{executables}
and other types of files. Therefore, attackers embed malicious codes into 
documents to download and install the actual malicious payload instead of
sending malicious payload directly or using vulnerability exploits. 
Moreover, we also found that financial matters (e.g., purchase orders and
invoices) are still the most common subject seen in Business Email Compromise
(BEC) scams that aim to trick \revision{employees}.
Finally, based on our analysis on the timestamps of captured malware samples, we
found that 93\% of the malware samples were delivered on weekdays. Our further
analysis also showed that
while the malware samples that require user interaction such as macro-based
malware samples have been captured during the working hours of the \revision{employees},
the massive malware attacks are triggered during the off-times of the 
\revision{employees}  to be able to silently spread over the networks.


\end{abstract}

\section{Introduction} \label{sec:intro}

Despite its ever-evolving nature, malware still is the most frequently
encountered cyber threat in the world~\cite{ENISA}, severely
impacting both enterprise and home networks. The damage caused may vary
depending on the type of malware and digital assets accessible on the victim's network.
Generally, a user has one or a few devices connected to home
network, while the number of systems connected to an enterprise network can vary
from hundreds to thousands with a variety of security policies in place.
This complexity of the enterprise networks brings new challenges for
securing valuable assets on such networks.

Reports~\cite{microsoft1} show that enterprise and home users are exposed to
different types of attacks because of their distinct day-to-day usage patterns.
Therefore, attacks may also differ from each other in several ways. First, since
attacks on enterprise networks can be very profitable, 
\abbas{attackers can be extra motivated to use more sophisticated, advanced, and persistent methods (i.e., targeted attacks).}
Second, since  enterprises prefer 
defense-in-depth approaches, which pose some restrictions on the use of the Internet
and email for personal purposes, enterprise users may face less number of
attacks but a wider variety of malware threats than the personal computer
users~\cite{microsoft1}. Last but not least, as the attack surface is much
larger on enterprise networks, with one insecure vector on the network (e.g.,
a misconfigured router),  attackers can access the data of multiple users, and
potentially stay undetected for longer periods of time. 


Even though malware detection is a well-studied topic in the literature, only a
few works~\cite{oprea1,oprea2,oprea3,oprea4} have focused on malware samples
encountered in real-world enterprises. These studies analyze security logs in
order to extract intelligence on malware discovered in a specific
enterprise. In another work~\cite{le2014look}, suspicious emails and malicious
attachments were examined. Compared to these studies, in this work,
we analyze the samples captured on-site inside two different enterprises, i.e.,
not only email attachments but also file downloads. Our work aims to shed some
light on what kind of malware is seen in typical, high-profile enterprises
today, what the infection vectors look like, and what trends do attacks follow. 

In this study, we have access to a dataset of $\approx3.6$
million samples collected from two enterprises from 2017 to early 2018 (we call
Organization A and Organization B to avoid disclosing their identity and security weaknesses).
Particularly, all the file downloads and email attachments of the employers from
two high-profile, global enterprises are collected and analyzed. Among all the
samples, only 2,942 of them have been detected as malicious, and among malicious
samples, the dataset includes 
\revision{122} 
unique samples that have never been seen in
the wild (i.e., not on VirusTotal (VT)), even as of the writing of this paper\footnote{\revision{July 2018}}. 

Moreover, this dataset has several unique features that other studies in
literature do not have: 

\begin{enumerate}

    \item \emph{The samples are captured on-site inside companies.} Previous
studies in the literature of malware research use datasets of malware that were
captured in the wild, or shared by AV companies. Hence, the malware being
analyzed does not have any context for how and when the infection took place.
However, our dataset has been captured through the sensors deployed on the
real-world networks of two enterprises.
    
    \item \emph{Both the behavioral analysis and Virus Total (VT) reports have
been obtained during both the time of capture (i.e., just after the sample has
been captured).} To the best of our knowledge, we are not aware of any study
using such a unique dataset. 
    
    \item \emph{The samples are analyzed using an advanced behavioral analysis
module that we have access to\footnote{We explain the details of the analysis
module in Section~\ref{subsec:analysis}.}} This module is able \revision{to} detect \abbas{2,920}
different malicious activities of the malicious samples and the list is always
updated with newly discovered malicious behaviors.    

\end{enumerate}

We leverage this dataset to perform the empirical analysis of malware samples
collected from two organizations. We characterize our analysis under five
categories. 

\begin{itemize}

    \item \textbf{Overall characteristics analysis:} In this, we analyze the
overall statistics of both benign and malicious files in the dataset to
understand the characteristics of files received and sent by enterprise
employers during their daily routines. 

    \item \textbf{Threat vector analysis:} In this, we analyze the document
types of malware used as a threat vector to infect the enterprise networks in
our study.  

    \item \textbf{Social-engineering analysis:} In this, we analyze the file
names of the malware, and the content of the malware instances to understand how
users are motivated to click on the malicious artifact. 

    \item \textbf{\revision{Vulnerability} analysis:} In this, we analyze the samples that have
been labeled with a CVE number in terms of their distribution over time. 

    \item \textbf{Time-series analysis:} In this, we analyze the distribution of
malicious samples over time to understand the logic behind the time management
of attackers. 

\end{itemize}

\begin{enumerate}

    \item On average, one out of two malicious samples in our dataset were not
detected by AVs on VT while almost one out of five malicious samples were not
found on VT during the time of capture of the sample.

    \item Documents are the most frequently used file types in both enterprises
with the frequency of 72\% and 36\%, respectively for Enterprise-A and
Enterprise-B. \revision{However, our further threat vector analysis showed that the file type distribution of malicious files is same as all files. While the most malicious file types are 
document\abbas{s}
, executable and jar
are the most common two file types used in malicious samples received in Enterprise B.}


    \item Our threat vector analysis showed that 34\% of all malware samples are
received in the format of \textit{jar} and those malware samples are labeled as
being part of massive phishing email campaigns by both AVs, and the dynamic
analysis module we had access to. 

    \item Our social-engineering analysis showed that 51\% of the malicious
documents are related to a financial matter (e.g., purchase order, invoice)
noting that financial subjects are the most used subjects in BEC scams. However,
contrary to reports~\cite{ENISA}, we also found that 23\% of all document-based
malware samples are organizational-looking (e.g., attached CV) files.

    \item Our \revision{vulnerability} 
    analysis revealed that 80\% of the malware samples
exploiting any CVE vulnerability are using the CVEs released in the year of
2017. This shows that attackers follow recent exploits, and use them more than
they use older exploits. We also verified other works~\cite{le2014look} that all
of the samples utilizing an exploit has been captured after their publish date.

    \item Finally, our time-series analysis revealed that as one would expect,
the number of received malware during the work hours is a lot more than those
captured during off times -- assuming that the employees work from 8 am to 5 pm
during weekdays. 
In contrast, there have been reports~\cite{weekend-stat,wannacry} that have shown that some large-scale, non-human interaction requiring 
attacks occurred during the weekend. 

\end{enumerate}

\section{Scope, Dataset, \& Privacy}

In this section, we explain the scope of the paper and the characteristics of our dataset.

\subsection{Scope}

Enterprise malware is not well-studied in literature because gaining access to
malware samples captured by enterprises is generally difficult.  However,
understanding the nature of the threats that enterprise users have been exposed
to during their daily works is important as such threats may result in
catastrophic outcomes.

Compared to home users, as a part of their daily work, the enterprise users
receive and send a lot more files, especially documents. Email is the most
common way of communicating and transferring files, which makes it also the most
common threat vector~\cite{dbir_verizon,malware_byte}.  However, other than
allowing us to have access to email attachments and file downloads, our dataset
does not include information related to the infection vector. That is, we do not
have access to the contents of the emails, email headers (e.g., from, to,
subject), or security logs inside the enterprise. Clearly, such information
would greatly help to an analysis like ours in this work, but such information
is typically difficult to acquire because of privacy concerns. 

\subsection{Dataset}

Every file downloaded, including email attachments from two enterprises during one year, have been captured through the sensors deployed at the organizations. \revision{\abbas{Sensors scan the incoming and outgoing network traffic,} the traffic within the network, as well as the host activity on the network. As the sensors are directly installed on end-users' systems, \abbas{they have} access to the unencrypted payload.}
Samples are sent to the back-end of the security company that we gained access to, and all the samples are analyzed in an isolated sandbox during the time of the capture.

\begin{table}[t]
\centering
    \begin{tabular}{ccccc}
    \hline
    \textbf{Organization} &  \textbf{Time interval} & \textbf{\revision{\# Samples}}     &  \textbf{Malicious samples (\revision{\%})}  \\ \hline \hline
    
    A            & Jan 2017 -  Feb 2018 & \abbas{3,192,452} & 243  (\revision{0.008 \%})                                                                         \\
    B            & Feb 2017 -  Jan 2018 & \abbas{463,476}   & \abbas{2,699} (\revision{0.582 \%})                                    \\
        Total            & Jan 2017 - Feb 2018& \abbas{3,655,928} & \abbas{2,942}  (\revision{0.081 \%})                \\\hline
    \end{tabular}
    \vspace{10pt}
    \caption {Summary of our dataset collected from two different organizations.}
    \label{table:summary}
\end{table}

\revision{\abbas{Our dataset includes reports of both benign and malicious samples}, which are indexed based on their hashes and as well as the raw malicious files. Moreover, we have access to reports generated at different times. Both behavioral analysis results and VT results of all files have been generated at the instant of the capture. Moreover, since VT results may change over time, we also checked the VT results during the time of the experiment.\footnote{\revision{July 2018}} Particularly,} the dataset includes the analysis result of $\approx3.65$ million samples (3.2M
from A and 450K from B), which have been captured from two organizations, namely
A and B\footnote{For privacy reasons, we do not disclose the company names.},
starting from \revision{January} 2017 to February 2018. The maliciousness occurrences of
samples collected from A is \revision{0.582 \%}, i.e., almost every 6 files
out of a thousand files an employee in the organization works on are malicious,
and it is much less than 1 in a thousand in organization B. As we mentioned
earlier, we also have the raw binaries of 
\revision{2940} 
malicious samples. In the
following sections, we analyze the characteristics of the malicious samples in
more detail.  Table~\ref{table:summary} is the summary of main characteristics
of our dataset.

\begin{enumerate}
    \item Behavioral analysis \revision{results}:
        \begin{enumerate}
        \item Metadata
        \begin{enumerate}
        \item Timestamp
        \item Hash (i.e., SHA1)
        \item File type
        \item Mime type
        \end{enumerate}
        \item The list of malicious behaviors 
        \end{enumerate}
        
    \item VT results
        \begin{enumerate}
        \item VT result (e.g., detection ratio, label) at the instant of the capture
        \item VT result during the time of the experiment\footnote{\revision{July 2018}}
        
        \end{enumerate}

    \item Malicious raw binaries

\end{enumerate}

\subsection{Privacy}

Note that even though we have the privilege of having a unique, real-world
attack datasets from two high-profile organizations, due to privacy policies,
our analysis has some limitations. In particular, we could not correlate the
captured data, and some features that are unique to the enterprises (e.g., such
as the industrial sector that they are active in).

\section{Analysis of Samples}
\label{subsec:analysis}

In this section, we explain our analysis methods, results and labeling
procedure.  Particularly, we used two types of analysis results to label the
samples: An advanced  Dynamic Analysis (DA) module that was provided to us, and
VirusTotal (VT). 

\noindent \textbf{Dynamic Analysis module reports.} The DA module is an advanced
malware detection and analysis module that runs the samples in a sandbox and
monitors their behaviors. It is capable of running the sample in an appropriate
environment for different file types. For example, if the sample is an
executable or document file (e.g., word, pdf), it is directly run in the proper
OS (e.g., Windows) environment and its behaviors are monitored. However, if it
is, for example, an archive file (e.g., zip, jar), it will be decompressed
first, and then executed. In addition, if it is an HTML or URL type of sample,
it will be executed in an instrumented or emulated browser. \revision{A malware sample, sometimes, can \abbas{run inside} more than one environment. For example, a JavaScript file can be executed by loading it in a browser as well as run directly on the operating system. While the sample reveals malicious behavior in an environment, it may not reveal in another environment.}  
In total, \abbas{2,920} different malicious sub-behaviors
under 35 total categories (e.g., evasion, packer, macro, signature) are
extracted. \revision{A sample is tested against all these malicious behaviors. If the sample shows a particular malicious behavior, that specific behavior has been added to the report. The report includes all of the malicious behaviors that have been revealed by the sample.} A sample report that is generated after monitoring the behaviors of the sample
is given in Appendix~\ref{sec:appendix}.

After acquiring the reports of malicious behaviors and sub-behaviors, in order to classify the
sample, every malicious behavior category (e.g., evasion, packer) is converted
into a boolean value according to the detection of the malicious sub-behavior
from that category. After that, every value is multiplied with its unique weight
and summed. The final result is called a ``score''. If the score of a sample is
less than 30, it is labeled as benign. If it is larger than 70, the sample is
labeled as malicious. \abbas{Samples with a score between 30 and 70 are labeled as
suspicious.} \revision{Note that we improve this simple classification by re-labeling the samples using the strategy in Table~\ref{table:labeling}.}

\begin{table}[t]
\centering
\begin{tabular}{lllll}
& & \multicolumn{3}{c}{\textbf{VT positives}}\\ 
& & \multicolumn{3}{c}{}\\ 
 \multirow{2}{*}{\rotatebox{90}{\textbf{DA module}}} & &\multicolumn{1}{c}{\textbf{Malicious(>3)}}&\multicolumn{1}{c}{\textbf{Benign(<=3})}&\textbf{NotFound} \\ \cline{3-5}
 & \textbf{Malicious(>70)} & \multicolumn{1}{|l}{\circled{1}-Malicious (\abbas{2,685})} & \circled{2}-Suspicious (128) & \circled{3}-Malicious (127)  \\

& \textbf{Suspicious([30,70])} & \multicolumn{1}{|l}{\circled{7}-Malicious (128)}
&\circled{8}-Suspicious (449) & \circled{9}-Suspicious (92)\\

& \textbf{Benign(<30)} &\multicolumn{1}{|l}{\circled{4}-Suspicious (285)} &\circled{5}-Benign ($\sim$3.6 M) & \circled{6}-Benign (\abbas{20,628})\\

\end{tabular}
\vspace{10pt}
\caption {Ground truth labeling strategy. DA score has been obtained through the dynamic analysis module and VT positives is the number of AV detection of the given sample on VT. }
\label{table:labeling}
\vspace{-10pt}
\end{table}

\noindent \textbf{VirusTotal reports.} We also checked the analysis results of
the samples on VT. As our dataset is dominated by the samples labeled as benign
by our DA module, the number of benign samples with 0 score are $99.8\%$
($\approx3.65$\revision{M}) of all samples. Therefore, we randomly selected a
\revision{subset} 
of benign samples, and checked those samples on VT. We observed that none are on VT
-- hence, not detected by AVs. However, we also checked all other \abbas{19,867} unique
samples with any type of malicious behavior (i.e., score$>0$) detected by our
dynamic analysis on VT.  In order to avoid false positives of VT, we chose the
threshold detection number of 3~\revision{\cite{perdisci2008mcboost}}. That is, if a sample is detected by more than 3
AVs on VT, we say that the sample is labeled as malicious by VT. Otherwise, it
is labeled as benign by VT. \revision{}

\noindent \textbf{Ground truth.} In order to obtain a ground truth for the
labels of the samples in our dataset, we use both the reports generated by the
DA module and VT. Even though there is a consensus on some of the files, there
are also inconsistencies between the DA module and VT.  We follow the strategy
in Table~\ref{table:labeling} in order to re-label the samples. In particular,
if labels from both the dynamic analysis engine and VT match (\circled{1} and
\circled{5}), the sample is labeled with the result of both tests, while if they
contradict each other (\circled{2} and \circled{4}), we label them as
suspicious.  Moreover, not all of the samples were found on VT, where we had
only DA module reports. For those  (\circled{3} and \circled{6}), if it is not
labeled suspicious (i.e., the score is not in $[30,70]$) by the DA module, we
labeled that sample with the result of that one report.  Finally, if the sample
is labeled by the DA module as suspicious, and found malicious by VT
(\circled{7}), we label it as malicious. However, if there is no consensus
between the DA module and VT, we label it as suspicious. \circled{8} and
\circled{9} are labeled as suspicious because either there are not enough
reports, or the samples are not exhibiting enough malicious behavior. \revision{The analysis of suspicious files has been left as a future work as the scope of this paper is to characterize the malicious files.}

At the end of this labeling procedure, we classified every sample as either
malicious, suspicious, or benign. In total, we have $\approx3.6 M$ benign,
\abbas{3,767} suspicious, and \abbas{2,940} malicious samples. Note that in this
classification, we used the most recent reports generated by VT, where we have
also the VT reports generated during the time of capture.

\section{Results}
In this section, we present the results from a more detailed analysis and share
our findings and insights. First, we analyze the overall characteristics of
the dataset. Second, we analyze the file types of malicious samples to understand
the threat landscape and infection vectors used in attacks. Third, we
investigate the malicious documents in detail to understand the
social-engineering techniques used to trick users. Fourth, we study the exploits
and the corresponding CVEs found in our dataset. Finally, we compare the data
from two organizations and discover the commonalities and differences from the
time-series distribution of malware samples.

\subsection{\abbas{Overall Characteristics Analysis}}

\begin{table}[t]
\centering
    \begin{tabular}{|c|c|c|c|c|}
    \hline
    \textbf{Organization} &  \textbf{Class}  &  \textbf{AV label}     &  \textbf{\begin{tabular}[c]{@{}c@{}}AV detection\\during the \\ time of capture \end{tabular} } &  \textbf{\begin{tabular}[c]{@{}c@{}}As of analysis\\(7/24/2018)\end{tabular}} \\ \hline \hline
    A and B           & Malicious & Undetected   & \abbas{1,318} (47.7\%)                    &  121 (4.4\%)                   \\
    A  and   B        & Malicious & Unclassified & 514  (18.6\%)                   & 122 (4.4\%)                              \\
         
\hline
    \end{tabular}
    \vspace{10pt}
    \caption {Number of unique unclassified and undetected malware samples. \label{table:dist}}
    \vspace{-10pt}
\end{table}

\subsubsection{Unique Malicious Samples.} 
We obtained the hashes of \revision{2,940} 
malware samples. We observed 2,766 unique
samples in total. These samples were checked on VirusTotal (VT) both at the time
of capture and at the time of this analysis. We call a sample: (1)
\textit{unclassified} if it is not found on VT; (2) \textit{undetected} if it is
found on VT and detected by fewer than 3 anti-virus software (AV) on VT. We
found that, at the time of capture, a small fraction of the unique samples
already existed on VT and were detected by AVs;  47.7\% of them existed on VT
but were not detected by AVs (i.e., undetected); 18.6\% of them had never been
submitted to VT (i.e., unclassified). It is worth noting that AVs on VT are
regularly updated with signatures of newly discovered malware. 
We found that the numbers of undetected and unclassified samples drop
significantly months later at the time of this analysis. \revision{We also observed that most of those unclassified samples are in the format of the document.} This result underlines
the delayed detection by AVs and thus the unsuitability of AVs for immediate
detection of malware attacks at their onset. 
Table~\ref{table:dist} shows the sample counts and percentages of undetected and
unclassified samples at the two different times. 

\noindent \textbf{Summary of findings-1:} 
As shown in Table~\ref{table:dist}, \textit{at the capture time, almost one out
of two malicious samples (1,318/2,763) in our dataset were not detected by AVs;
almost one out of five malicious samples (514/2,763) had not been submitted to
VT.} This shows the ineffectiveness of the AVs in real-time malware detection.
However, AVs can still be useful as the first line of defense in the
defense-in-depth solutions deployed in enterprises, in order to quickly filter
out the previously discovered/reported malware samples. Moreover, we also found
that AVs evolve over time by adding more samples to their database. The
percentage of both undetected and unclassified samples dropped to $4.4\%$ at the
time of this analysis (i.e., months after the initial malware captures).
However, there were still 121 unique malicious samples undetected by AVs and 122 \revision{unique}
samples never submitted to VT at that time. \revision{Note that unclassified and undetected files refer to different files, so in total, we can interpret it as 243 files can not be detected by AVs. Moreover, we also note that as we did not perform analysis on historical VT reports.} 

\begin{figure}
    \centering
    \includegraphics[width=.8\columnwidth, trim= 0cm 0cm 0cm 0cm ]{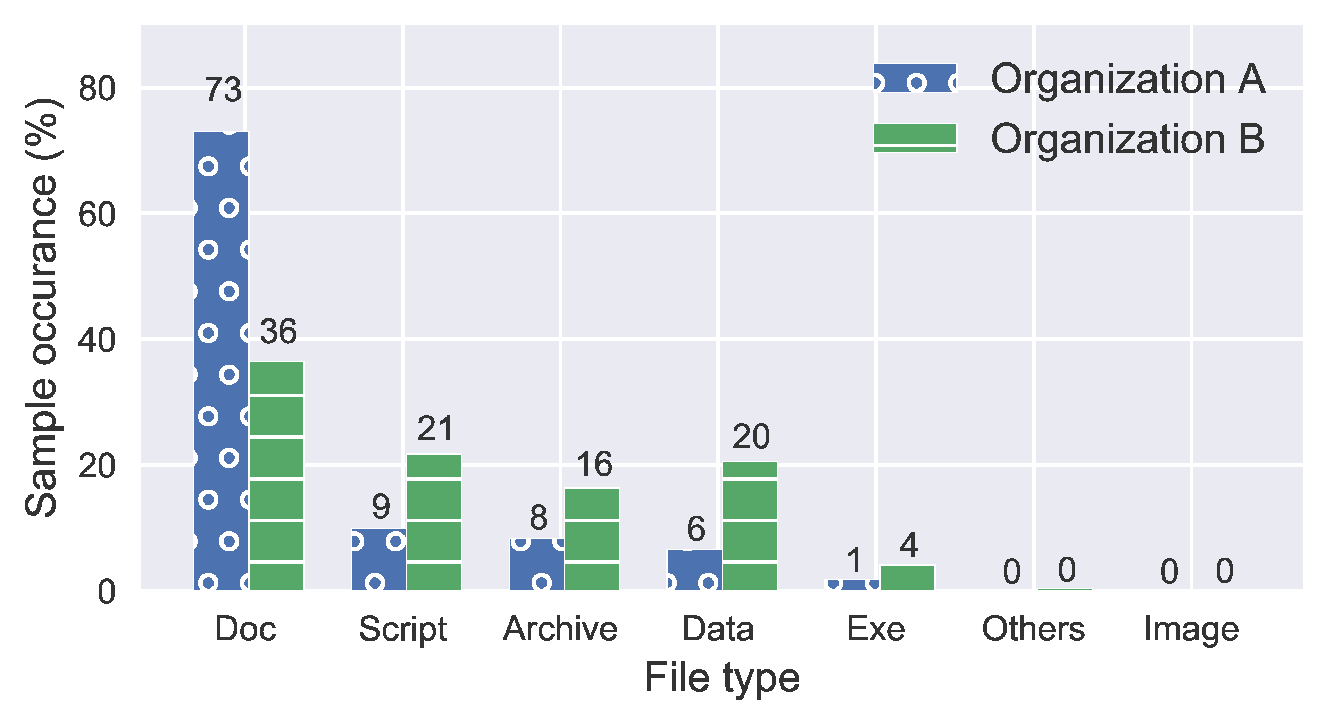}
     \caption{File type distribution among all samples ($\approx 3.65$ M samples), including both benign and malicious samples from Organization A and B. The figure shows that documents (e.g., doc, pdf) are the most common file types observed in enterprise networks whereas the executable file types are a lot less common. }
    \label{fig:file_type}
\end{figure}

\subsubsection{File type distribution.} The types of samples are detected and
reported by the DA module. Figure~\ref{fig:file_type} shows the distribution of
file types among all the samples in Organization A and B.
The most common file type used in both A and B is a document, as expected, with
frequencies of 73\% and 36\% for organizations A and B, respectively. Since the
number of benign samples is a lot more than the number of malicious samples, the
distribution is dominated by benign samples. However, as this is also known to
attackers, they can use documents to carry or hide a malicious payload in
order to bypass detection heuristics based on file types. Moreover, it is also
interesting to see that scripts such as JavaScript or PowerShell 
scripts are commonly used in these organizations for benign purposes. Although
we have no visibility into what exactly scripts are used for, we can reasonably
expect that may serve the purpose of automating workflows. We expect that the
attackers may increasingly utilize scripts in order to infect enterprise
computers and networks. Finally, we also observed that executable are much less
common (respectively 1\% 4\% for A and B) than document-based samples.
Therefore, considering that malicious payloads often exist in the form
executables, it is more likely for executables propagated in company networks,
especially from untrusted sources, to be malicious than documents.

\begin{figure}
    \centering
    \includegraphics[width=.8\columnwidth, trim= 0cm 0cm 0cm 0cm ]{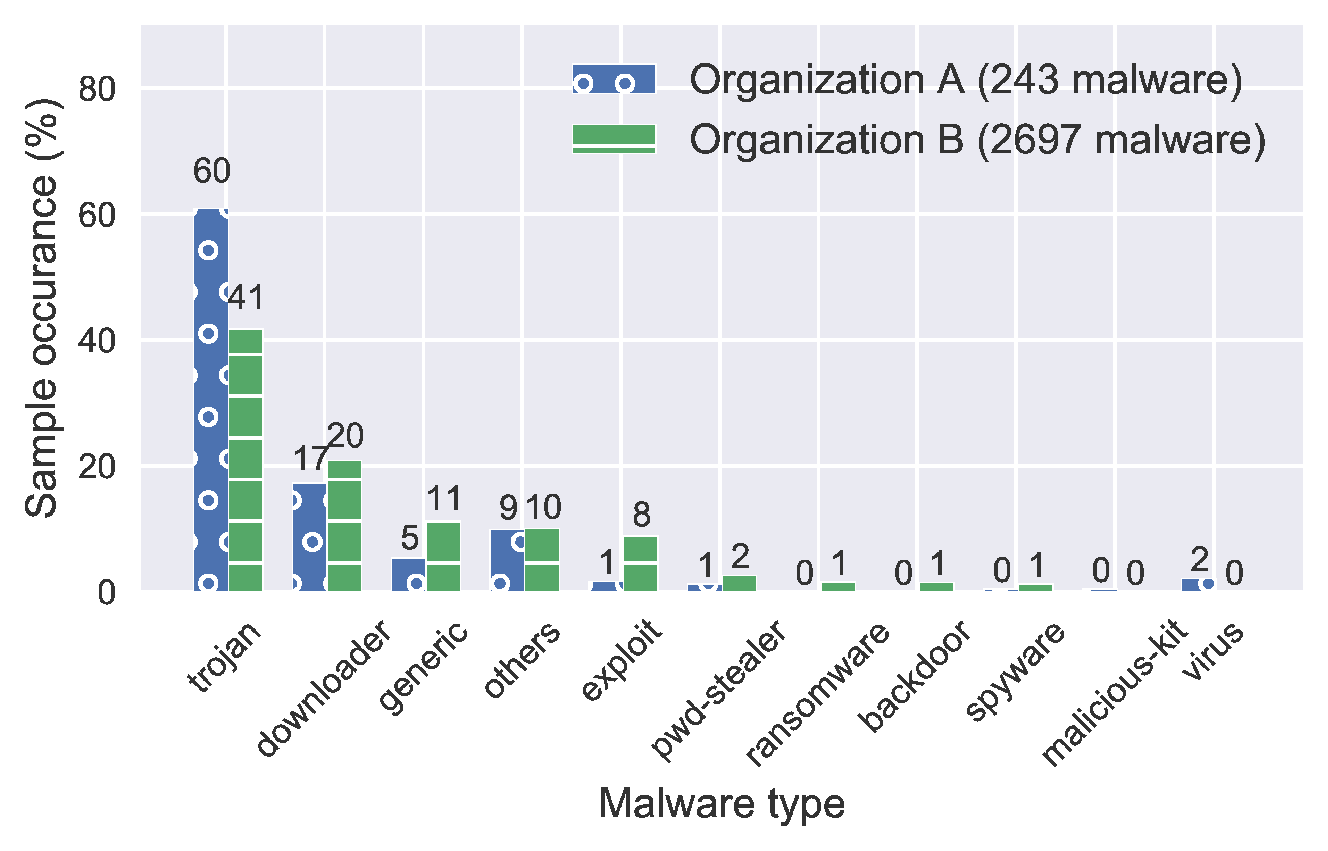}
     \caption{Top-10 malware types in our dataset, where trojan is the most common malware type for both A and B. Moreover, downloaders are a lot more common than exploits, which shows that attackers prefers simpler methods like macro-based malwares over more advanced 
     \revision{vulnerabilities}.}
    \label{fig:av_label}
\end{figure}

\subsubsection{Malware type distribution.} In order to better understand the
threat landscape in organizations, we also analyzed the types of malware samples
in our dataset. 
The distribution of malware types observed in A and B is illustrated in
Figure~\ref{fig:av_label}. 

We used the malware labels reported by the DA module. If a sample was undetected
and unclassified by the DA module, we fetched the most recent labels from VT. We
note that a sample often has different labels assigned by different AVs on VT
and many of them tend to use overly generic labels (e.g., trojan). We picked
more specific and descriptive labels, such as those from Microsoft and
Kaspersky. 


\noindent \textbf{Summary of findings-2:}  The top-10 malware types in our
dataset are shown in Figure~\ref{fig:av_label}. The most common malware type in
our dataset is trojan for both A and B. Trojan is a generic name and mostly used
for labeling samples that are not associated with any malware family or do not
contain enough malware family information. Therefore, as expected, the number of
trojan samples is more than other types in both A and B. In the second place, we
see the downloaders, counting for 17\% and 20\% of malware found in A and B,
respectively. The downloaders are usually embedded to documents and infect the
system by downloading and installing more malware at a later stage of
infection. This hidden malware is more difficult to catch due to their limited
and conditional exposure. Therefore, the methods based only on static analysis
are not able to extract these samples, echoing the need for combining both
static and dynamic analysis for malware discovery.  
\revision{The fourth in both Organization A and B is exploit, meaning that samples showing the behavior of a publicly disclosed CVE. \textit{As shown in Figure~\ref{fig:av_label} compared to downloaders and trojans, we observe that both organizations receive less number of exploits}.}
This result matches the findings
reported by some security companies~\cite{istr_2017,malware_byte}. It shows that
the attackers are more often using simple methods (e.g., embedding malware using
document macros) than employing advanced \revision{vulnerabilities.}
This is because
the former is much less complex to carry out than the latter while still
yielding good results on unwary or security-unconscious users in enterprises. 

\begin{figure}
    \centering
    \includegraphics[width=.8\columnwidth, trim= 0cm 0cm 0cm 0cm ]{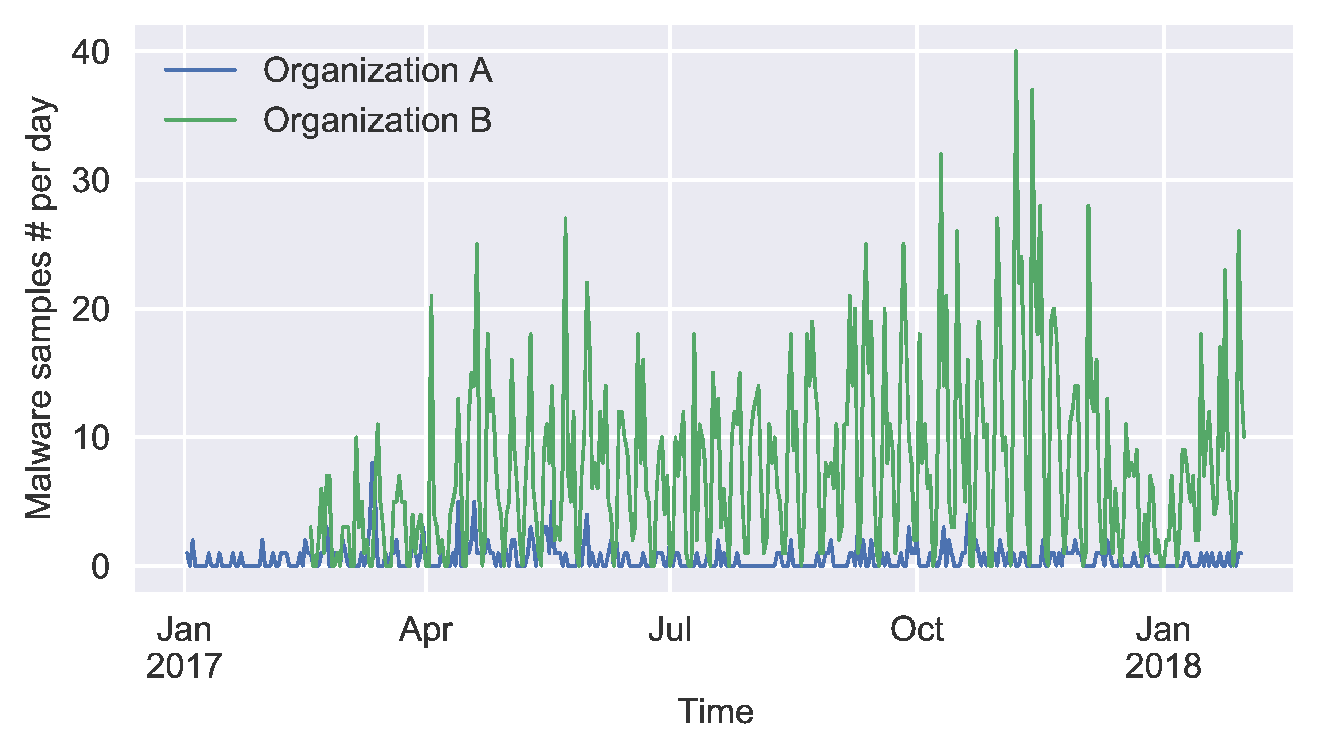}
     \caption{The number of malicious samples captured per day. On average, A received 0.6 malicious and 8K benign samples per day while B received 7.7 malicious and 1300 benign samples. }
    \label{fig:malware-per-day}

\end{figure}

\subsubsection{Average malware counts per day.}
We analyzed how the capture/appearance of malware varies over time and if it exhibits any patterns. 
Figure~\ref{fig:malware-per-day} shows the number of malicious samples captured
per day in our dataset during the course of a year (Jan. 2017 to Jan. 2018). 

\noindent \textbf{Summary of findings-3.} According to
 Figure~\ref{fig:malware-per-day}, the number of malicious samples per day
 varies in the range from 0 to 8 for Organization A throughout the whole period
 whereas the same number fluctuates significantly and topped at 40 per day for
 Organization B. \emph{On average, A received 0.6 malicious and 8K benign
 samples (downloads and email attachments) per day while B received 7.7
 malicious and \abbas{1,300} benign samples per day.}
B has seen 80 times more malicious samples than A. This discrepancy indicates
that the risks of attacks and malware infection can vary a lot across different
organizations and industry sectors, revealing attacks being driven by nature
and potential value of the target businesses. 
Moreover, we tried to identify the cause to the spikes in
Figure~\ref{fig:malware-per-day}. But we could not find any major security
events or reports for those dates. We suspect that the spikes were resulted from
some hidden attacks targeting that particular organization or sector.

\subsection{Threat Vector Analysis} 
\begin{figure}
    \centering
    \includegraphics[width=.8\columnwidth, trim= 0cm 0cm 0cm 0cm ]{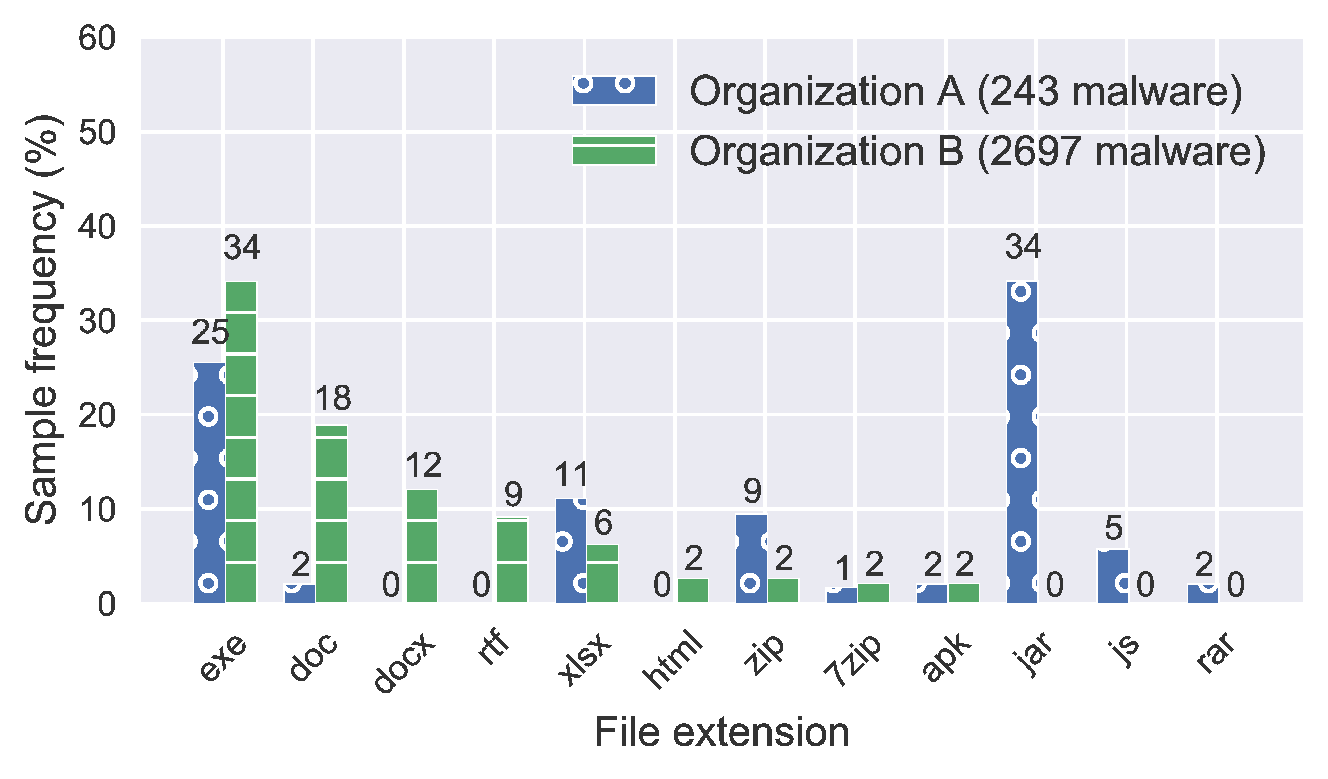}
     \caption{Most malicious file types in A and B, where 6 of them are common. }
    \label{fig:10_most_malicious}
\end{figure}

Attackers use not only executable files but also other types of  files such as
MS office documents (e.g. docx, xlsx) or archive files (e.g., jar, zip) to
spread malware. Using documents, the attackers can embed the malicious code and
run within a document itself. This embedding can happen in the form of, for
instance, macros in MS Office documents and JavaScript (JS) in pdf files. Since
MS Office 2016, macros have been disabled by default in documents
~\cite{macro_disabled}. Similarly, JS code has been disabled by default in PDF
readers. Therefore, unlike executables, the malicious code does not execute in
the first place when a user opens the document. Instead, the user is asked to
"enable active content". To adapt to such security countermeasures, attackers
now try to convince users to enable the macros and scripts using
social-engineering techniques, e.g., showing images with fake warnings for users
to "Enable content". When accompanied by very convincing messages or emails,
users may easily fall for these tricks. Once the macros have been enabled, the
embedded malicious code is triggered, which goes on to download and run the
actual malware or full malicious payload. 
Moreover, Java (jar) archive files have been widely used and have seen a surge
in recent malicious email campaigns~\cite{jar_campains}, where the malicious
payload is compressed (zip or rar) and attached to the email. This file format is
preferred by the attackers as it is relatively less-known file type for
malicious files. Plus, it benefits from the cross-platform nature of Java. In
addition to macros and JS, we also observed other types of scripts. For example,
PowerShell scripts have been used to infect Windows PC. In
Figure~\ref{fig:10_most_malicious}, we plot the 10 most malicious file types
used in A and B, among which 6 are common between A and B.




\noindent \textbf{Summary of findings-4.} 
Figure~\ref{fig:file_type} shows that 73\% and 36\% of the files observed on the
company networks are document-based files, compared to only 1\% and 4\% for
executable files, for A and B, respectively. 
We found that Java archive files are the most common file type used in malware
samples targeting B while executable files are the most common for A. Both types
of files require manual actions to be triggered. Malware written in Java can run
on any platform that has JVM (Java Virtual Machine) installed, i.e., does not
require the knowledge of the system that the victim is using. Therefore, they
are highly preferred for phishing email campaigns. We also saw the samples of
commercial RAT (Remote Access Trojan), such as Jrat, Adwind, Jaraut. \textit{For
B, we observed that 34\% of all malware samples are jar files and we also saw
that all of the Java malware samples are labeled as part of \revision{phishing}
email
campaigns by AVs. Therefore, one can reasonably suspect that the list of emails
of A may have been leaked to attackers.} 
Moreover, \textit{our results confirm the findings of the other
reports~\cite{istr_2017,malware_byte} that  
document-based malicious files are still highly preferred by attackers as they
are often considered safe by the average users.} 



\begin{figure}
    \centering
    \includegraphics[width=.8\columnwidth, trim= 0cm 0cm 0cm 0cm ]{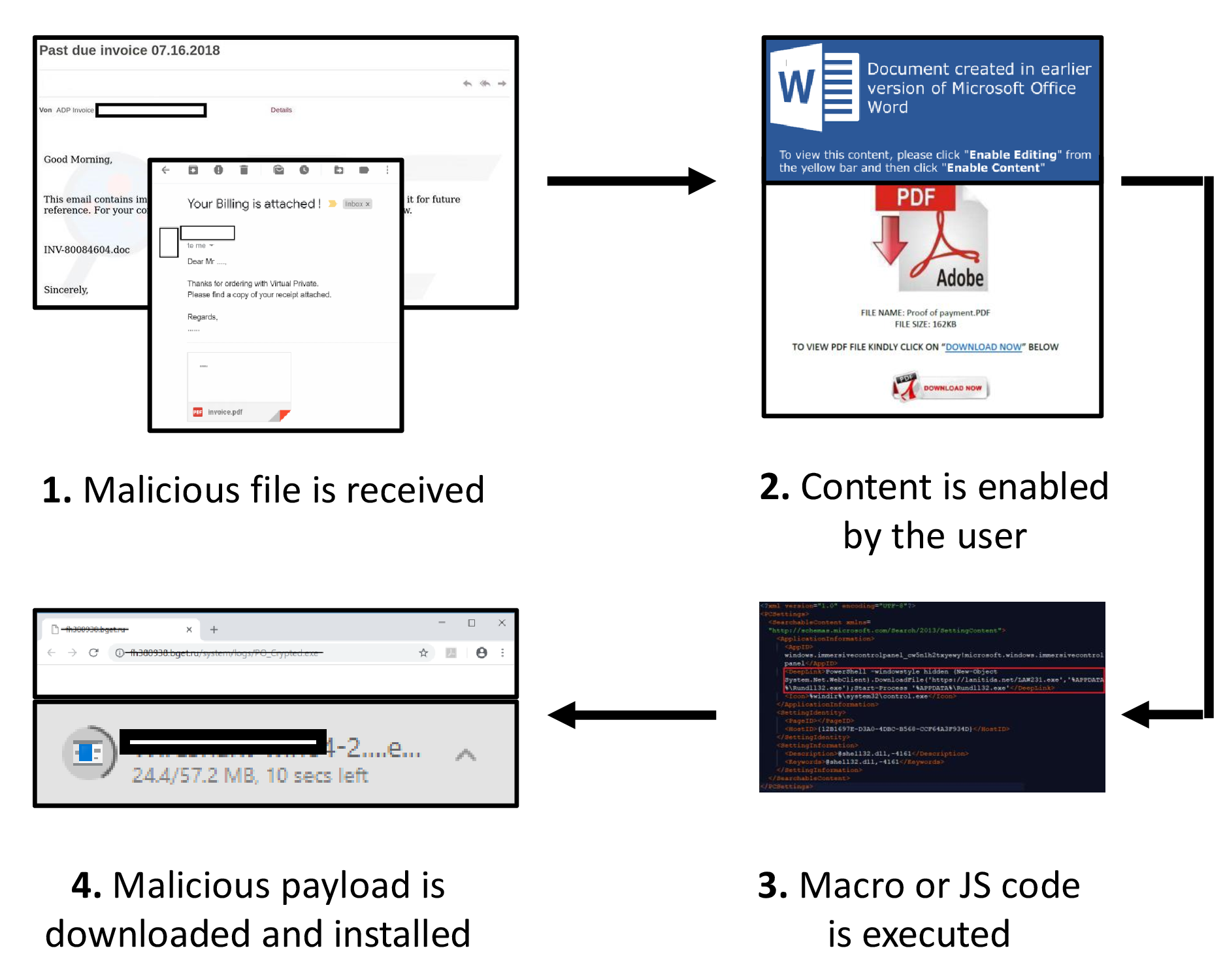}
     \caption{Infection chain of a sample document-based malware.}
    \label{fig:chain}
\end{figure}

\subsection{Social-engineering analysis}

According to recent reports published by Microsoft~\cite{microsoft1}, there is a
recent transition from exploits to macro-based malware to infect endpoints. In
our dataset, we observed 1,370 malicious documents while there is only 158
samples \revision{associated with a publicly known vulnerability.} 
Even though macros have been disabled by
default since 2016, it continues to be a common threat vector for massive
pishing campaigns like Locky, Cerber~\cite{pishing_camps}. This still poses
threats mainly because people fall victim to social engineering tricks, and in
turn, grant permissions to malicious samples. Therefore, it is important to
understand and prevent social-engineering techniques used by the attackers.


What happens if a user is tricked to enable the macros or JavaScript in pdf
documents? In order to find out, we also run the samples in an isolated
environment and observed their behaviors. Figure~\ref{fig:chain} shows the
infection chain of document-based malware samples. Malicious documents are
usually received as an attachment to emails, which direct users to enable dynamic content or scripts. When a user does so, the code runs, downloads the
actual malicious payload, and then installs it on the victim's system.

\noindent \textbf{Method.} In order to understand the social-engineering
techniques used by attackers, we performed the following analysis. First, we
analyzed the file names of the samples when received by the victim. Since the
samples in our dataset were renamed using their hash values, we fetched the
original filenames of the documents from the VirusTotal's database. Second, we
also analyzed the subject of the document as inferred from file names and actual
file content. Based on inferred subjects, we categorized all the documents into
several categories. We observed that some of the samples only include "content
enable" images and do not have a meaningful file name. We categorized them as
No-content files.

\begin{figure}
    \centering
    \includegraphics[width=.8\columnwidth, trim= 0cm 0cm 0cm 0cm ]{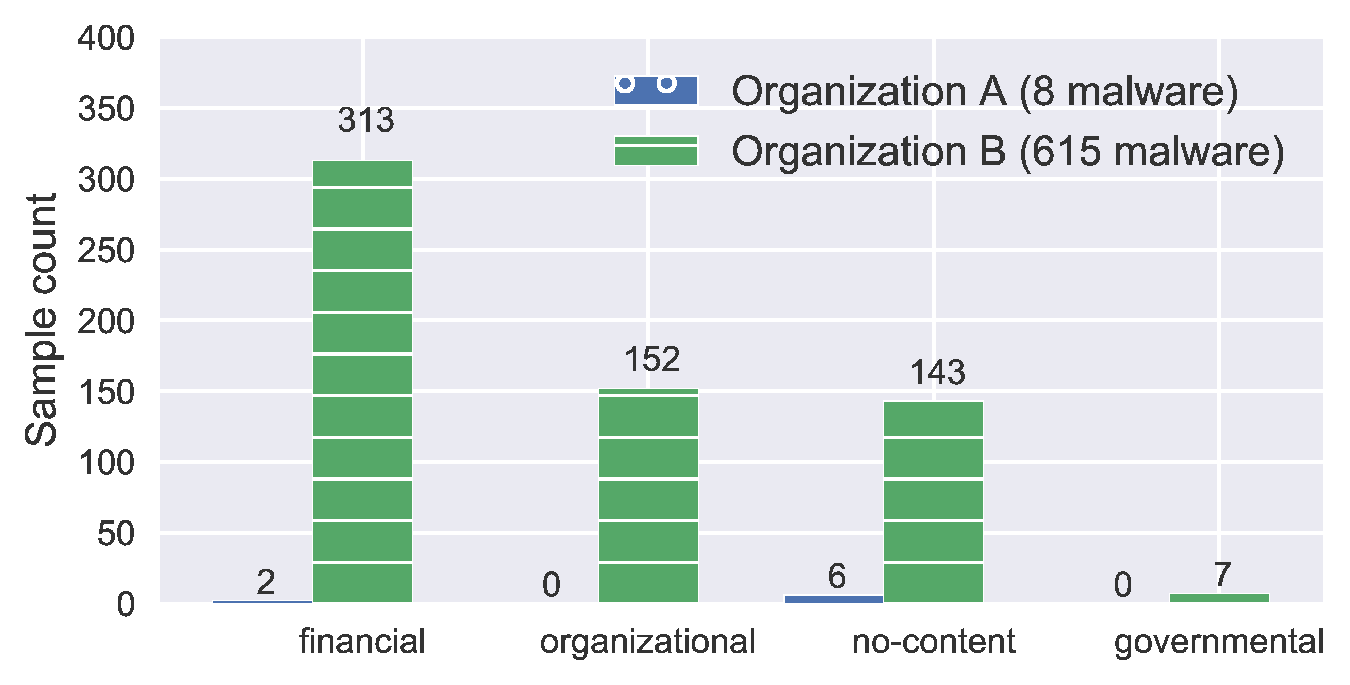}
     \caption{Distribution of the subject in malicious documents. \revision{\textit{51\% (313/615) of the malicious documents Organization B received have been shown as related to a financial matter, while it is hard to comment on Organization A, as it has a very limited number of samples (36 samples.)}}
     }
    \label{fig:mal_subject}
\end{figure}

\begin{figure}
    \centering
    \includegraphics[width=.8\columnwidth, trim= 0cm 0cm 0cm 0cm ]{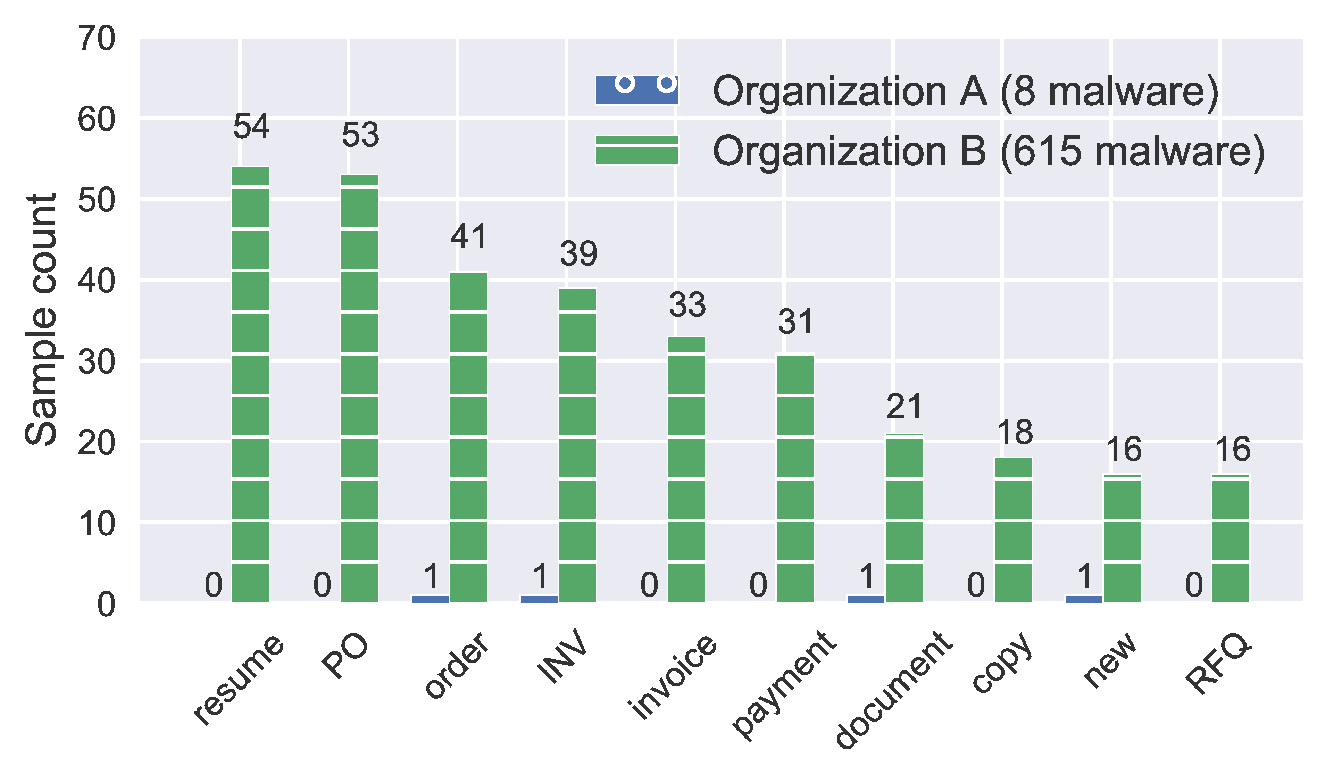}
     \caption{The top 10 keywords used in file name of the document-based malwares.}
    \label{fig:mal_names}
\end{figure}

\noindent \textbf{Summary of findings-5.} The subject distribution is shown in
Figure~\ref{fig:mal_subject}. \textit{51\% (313/615) of the malicious documents
Organization B received have been shown as related to a financial matter, which
is similar to the results reported by Symantec~\cite{istr_2018} showing that
financial subjects are the most used in business email compromise (BEC) scams.
However, unreported in \cite{istr_2017}, we found that 23\% of all
document-based malware pretends to be usual  business files of various kinds.}
For example, emails related procurement orders and resumes are highly common in
the malicious documents found in B.  In order to better understand the
techniques used by attackers, we also analyzed the file subject in more details
and plotted the counts for each subject word (Figure~\ref{fig:mal_names}). The
most commonly used keyword is "resume", accompanied by a name (e.g.,
"Rebecca-Resume.doc"). Other commonly used phrases are mostly related to 
finance such as "order", "invoice", or "payment". Moreover, their acronyms like
"PO", "RTQ", "INV" etc. are also mostly preferred to trick the victim. 


\subsection{Vulnerability Analysis} 
Newly discovered and reported vulnerabilities in software
are assigned an ID,
called CVE, as a uniform reference among vendors and security researchers. 
If AVs detect that a malicious sample exploits a vulnerability, it labels the
sample with its publicly known CVE ID. In this section, we share the details
about the vulnerabilities found in the samples and their characteristics. 

\noindent \textbf{Method.} In order to tag the malware samples with a CVE
identifier, we used three sources. The first one is our DA module. It matches
the vulnerability behavior with the malware's behavior \revision{i.e., if the particular behavior is observed, the sample is labeled with the given CVE instead of malware type related to that behavior.} Second, we used the
labels provided by Microsoft and Kaspersky on VT. We observed 158 samples in
total that use at least one vulnerability with a CVE. We observed a discrepancy
between the samples labeled by them. For instance, five samples are tagged as the
exploit of CVE-2014-1761 by Kaspersky and as CVE-2012-0158 by Microsoft. We
counted those samples twice, which does not affect the overall results much due
to the small numbers of such cases. \revision{}

\begin{table*}
\centering
\resizebox{\textwidth}{!}{%
    \begin{tabular}{lccccccc}
    \hline
    \textbf{CVE ID} & \textbf{Publish Date} & \textbf{First seen on VT} & \textbf{Capture Time} & \textbf{Count} & \textbf{Affected product}     & \textbf{Vulnerability Type} &  \textbf{AV detection ratio}   \\ \hline \hline
    2010-3333   & 2010-11-09 & 2017-05-09 & 2017-05-09&2  & MS Office & Remote Code Execution &  $30/56$    \\
    2012-0158     & 2012-04-10 &2017-03-22&2017-03-22& 22 & MS Office&  Remote Code Execution & $32/58$ \\
    2014-1761    & 2014-03-25 & 2017-08-21 & 2017-08-21 & 5  & MS Office &  Memory corruption & $30/59$ \\
    2015-1641   &  2015-04-14 &2017-10-04&2017-10-10&5 & MS Office&     Memory corruption &$31/59$\\
    2015-2545   &2015-09-08 &2017-09-08 &2017-09-08&1 &  MS Office&    Remote Code Execution &$27/59$\\
    2017-0199   &2017-04-12& 2017-05-23 &2017-05-23&64  & MS Office & Remote Code Execution & $31/58$    \\
    
    2017-8570  & 2017-07-11    &2017-08-30&2017-08-30&6  & MS Office & Remote Code Execution & $29/60$ \\
    2017-8759     &2017-09-12 &2017-09-14&2017-09-19&33  &  MS .NET Framework   &  Remote Code Execution & $27/59$     \\
    
    2017-11882   &2017-11-14 &2017-11-22&2017-11-22 & 23  & MS Office & Memory corruption & $33/59$    \\ \hline
     
    \end{tabular}
    }
    \vspace{10pt}
    \caption {List of CVEs in our dataset. \textit{Publish Date}, \textit{Affected product}, and \textit{Vulnerability Type} are taken from~\cite{cve_details}. \textit{First seen on VT} and \textit{Capture Time} are the respective dates for the first sample in our dataset. \textit{AV detection ratio} is the average detection ratio of all samples tagged with a specific \textit{CVE ID}.}
    \label{table:cve}
\end{table*}

\begin{figure}[h]
    \centering
    \includegraphics[width=\columnwidth, trim= 0cm 0cm 0cm 0cm ]{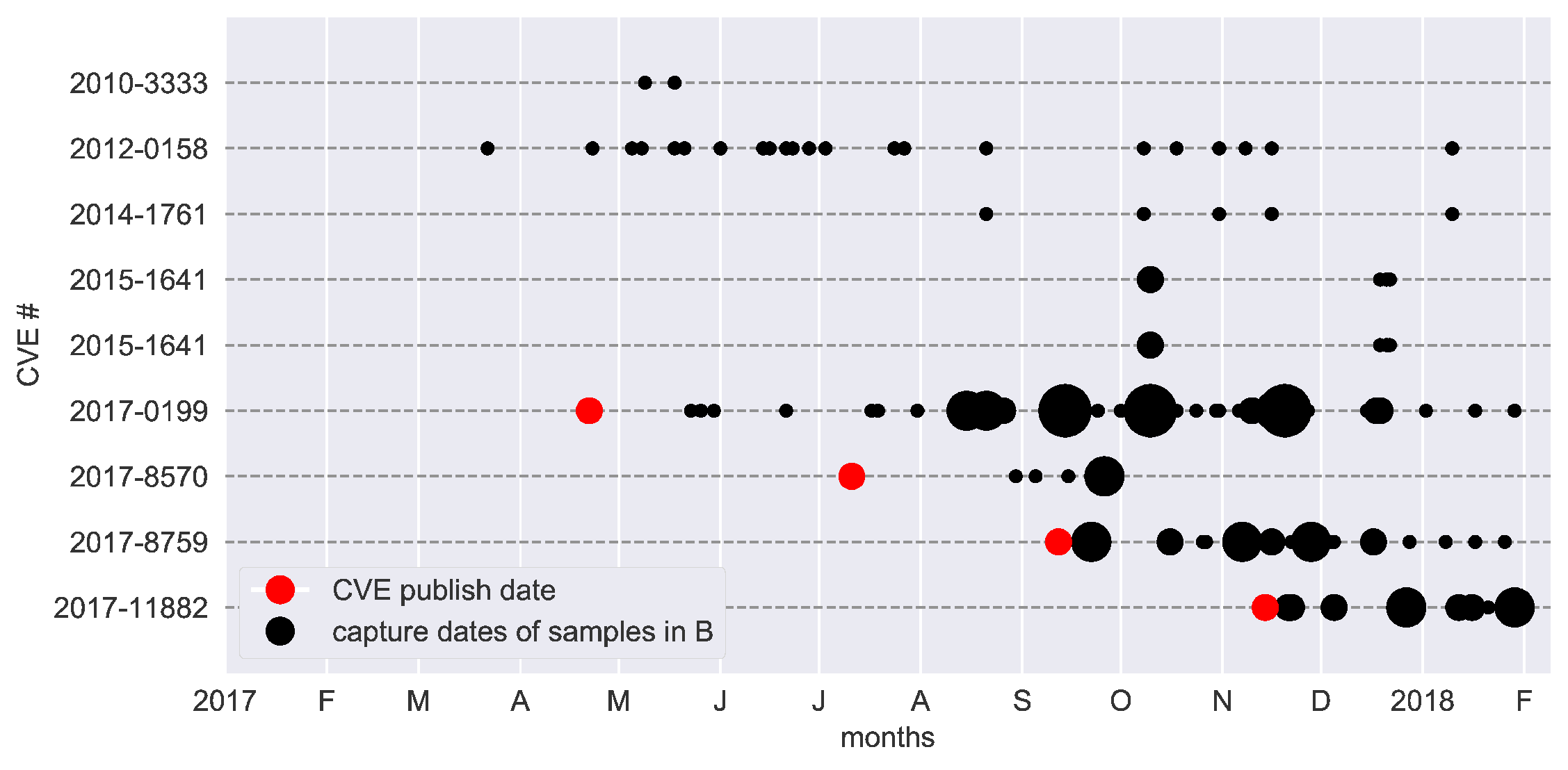}
     \caption{Distribution of CVEs in our dataset over time. The size of the circle is proportional with the sample count at that time interval.}
    \label{fig:cve-dist}
\end{figure}

\noindent \textbf{Summary of findings-6.} \revision{Table~\ref{table:cve} shows the list of nine CVEs found in our dataset. Publish date,  affected product, and vulnerability type are taken from [4]. First seen on VT and capture time are the respective dates for the first sample in our dataset. AV detection ratio is the average detection ratio of all samples tagged with a specific CVE ID.}  As shown in Table~\ref{table:cve}, \textit{80\% of the exploits are targeting
the CVEs released in the year of 2017. This shows that attackers tend to use
recent exploits for better results.} This is because many systems now
automatically patch known vulnerabilities within a short window (e.g., a few
months). The chances for a successful exploit is higher if more recent
vulnerabilities are targeted than older vulnerabilities. This also shows that
attackers are fast in following and leveraging new CVEs.

Moreover, we also saw that malware using CVE-2017-8759 appeared only two days
after its release date. The first sample exploiting CVE-2017-8570 was captured
50 days after its public release. In general, we observed that \textit{a sample
exploiting a vulnerability can be seen in the wild after a period of a few
months or even just days since the vulnerability disclosure.} Therefore, it is
important to patch vulnerabilities as soon as possible. On the other hand, when
we analyzed the samples for Organization B, we saw that \textit{on average, B
received the samples exploiting vulnerabilities three months after their
disclosure dates. }

Figure~\ref{fig:cve-dist} shows the distribution of exploits in our dataset over
time in terms of their captured dates and initial disclosure dates. \textit{In a
similar work~\cite{le2014look}, it was shown that none of the exploits were used
before their public disclosure. We also verify that all of the samples utilizing
an exploit were been captured after the CVE disclosure dates.}


\begin{figure}
    \centering
    \includegraphics[width=.8\columnwidth, trim= 0cm 0cm 0cm 0cm ]{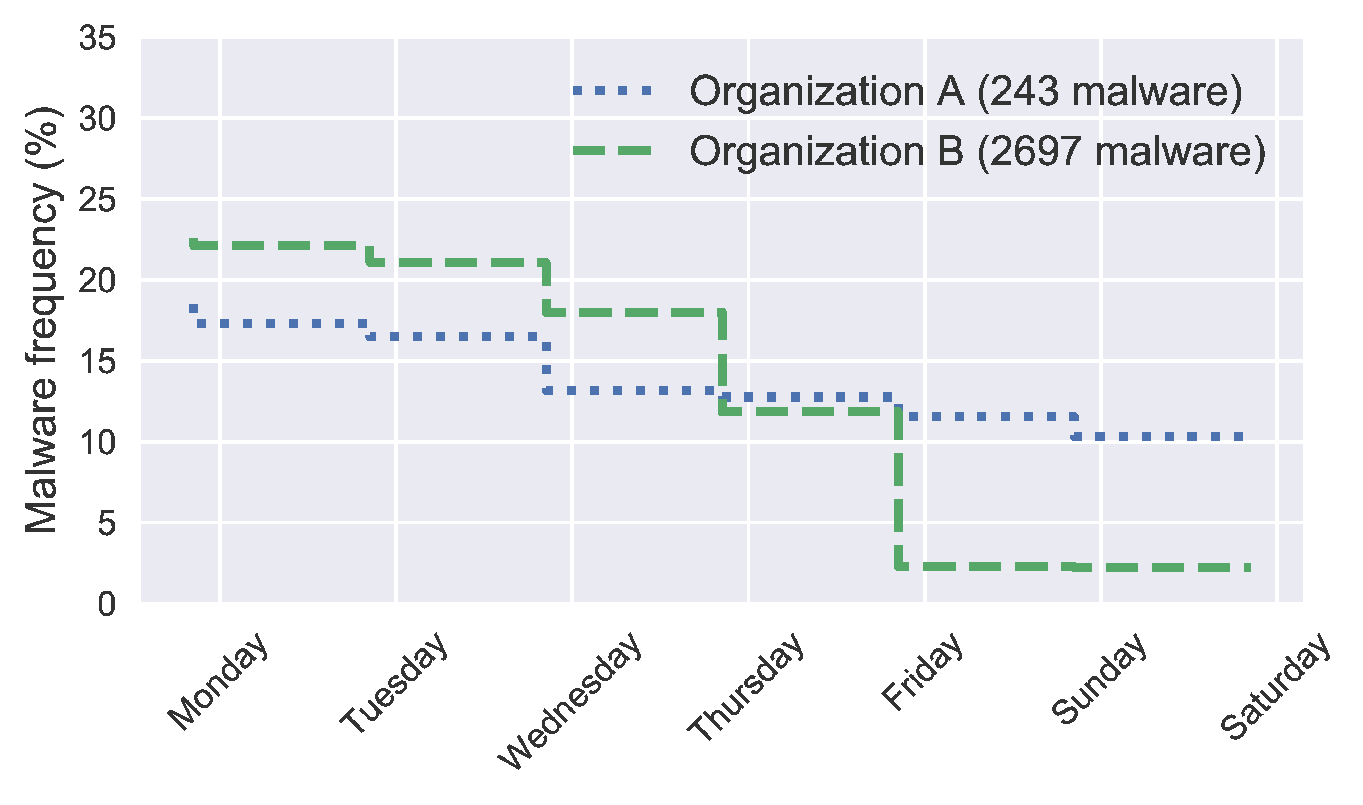}
     \caption{Number of malware received per the days of the week. }
    \label{fig:day}
\end{figure}

\begin{figure}
    \centering
    \includegraphics[width=.8\columnwidth, trim= 0cm 0cm 0cm 0cm ]{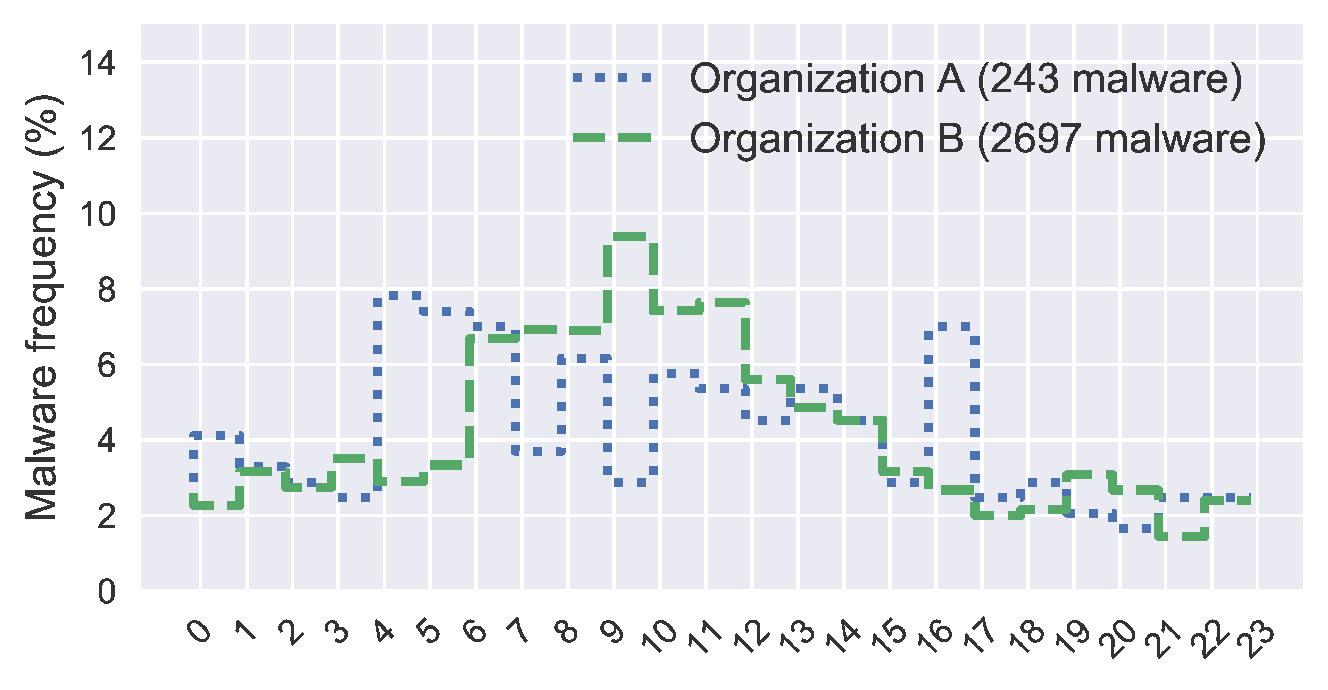}
     \caption{Number of malware received per hours of the day. }
    \label{fig:hour}
\end{figure}

\subsection{Time Series Analysis of Malicious Samples}
\revision{Attackers} 
use many different
techniques to trick users. Some of these tricks depend on proper timing. 
For example,
\revision{employees may receive the malicious samples at a certain time of the day, or 
specifically on some days.}
In this section, we analyze
these timing-related factors from the victim's perspective. We use the capture
time of the malicious samples in our time series analysis.

\noindent \textbf{Working hours vs. Off-times.} 
According to the reports~\cite{weekday-stat}, the largest number of security
threats are detected on weekdays, i.e., when employees are working on their computers. 
In order to verify if this pattern
exists in our dataset, we plot the distribution of malicious samples frequency
for each day of the week in Figure~\ref{fig:hour} and for each hour during the
day in Figure~\ref{fig:day}.

\noindent \textbf{Summary of findings-7.} 
Based on Figure~\ref{fig:day} and \ref{fig:hour}, \textit{the number of received
malware during the work hours is a lot bigger than that number for off-hours,
assuming employees work from 8 am to 5 pm on weekdays.} However, there are also
reports~\cite{weekend-stat} and some massive attacks~\cite{wannacry}
contradicting with our finding. Especially, the attacks utilizing \revision{a vulnerability} 
prefer weekends as they may want to spread over the network without
being detected. However, the malware that require to be enabled by the victim
are going to prefer working hours.

\section{Related Work}

Malware detection has been an active research area for years. There have been
numerous
studies~\cite{bayer2009scalable,hu2009large,tamersoy2014guilt,invernizzi2014nazca}
working with large-scale malware dataset with the sizes of the datasets changing from
a hundred thousand to millions and on different problems such as detection,
clustering, indexing, etc. In all these studies, the samples captured different
sources and were brought together for evaluating the proposed method, or
collected in the wild, where the target entity is unknown. However, in practice,
while it is known that home and enterprise users are known to have different
threat landscapes, it is not clear if different enterprises actually have been
targeted by different types of threats. In our study, we perform an analysis of
such a dataset and provide the characterization of malicious samples captured
from two different enterprises. 

\noindent \textbf{Enterprise malware detection.} In the literature, there are
only a few works~\cite{oprea1,oprea2,oprea3,oprea4,platon2019} about enterprise malware.
All of these studies analyze security logs in order to extract some intelligence
related to malware encounters occurred in a specific enterprise. In another
work~\cite{le2014look}, the authors analyze suspicious email collected from two
members of an NGO, where both the content of the emails, and malicious email
attachments were used for the analysis. Compared to the datasets used in these
studies, our dataset was collected from specific commercial enterprises from
2017 to early 2018 and the samples were analyzed during the capture time with
both a DA analysis module as well as VT. Note that to date, there have not been
many scientific works that have reported on what kind of malware high-profile
organizations are faced on a daily basis. In this paper, we aim to bridge that
gap.





%

\section{Conclusion}
\label{sec:conclusion}

In this work, we presented an analysis of malware samples captured from two
different enterprises from 2017 to early 2018. First, as one would expect, our
analysis on the combined dataset showed that only-AV solutions are not effective
in real-time defense because on average, 40\%  of the malware samples were
either not detected at all, or have never been seen in the wild yet during the
incident. Second, as employees in a typical enterprise work with more documents
than executables, attackers mostly use documents as an attack vector. Hence,
frameworks that allow the processing of documents in the cloud would provide
better protection against many such attacks. In 
\revision{our vulnerability analysis,}
we also found that attackers use recently disclosed CVEs more than older
disclosures. Additionally, after our social-engineering analysis, we also found
that financial issues are still the most common subject used in
social-engineering attacks against the enterprises that we analyzed.

\noindent \textbf{\abbas{Acknowledgments.} }
\abbas{The authors would like to thank the anonymous reviewers and our shepherd Dr. Yajin Zhou for their comments and suggestions, which significantly improve the quality and presentation of this paper. This work was partially supported by US National Science Foundation under the grant numbers NSF CNS-1514142, NSF CNS-1703454, NSF-CNS-1718116, NSF-CAREER-CNS-1453647, and ONR (CyberPhys).} 


\bibliographystyle{splncs04}
\bibliography{ref}

\newpage

\appendix


\section{A sample malicious behaviours report generated by the DA module}\label{sec:appendix}
The list of malicious behaviours extracted via the DA module from a given sample: 

\vspace{10pt}

\fbox{%
    \parbox{.9\textwidth}{\footnotesize
  "Anomaly: Found suspicious security descriptors for lowering the integrity level", \\
  "Autostart: Registering for autostart during Windows boot", \\
  "Evasion: Potentially malicious application/program", \\
  "Evasion: Potentially malicious application/program (MMX stalling code detected)", \\
  "Evasion: Targeting anti-analysis/reverse engineering", \\
  "Evasion: Trying to enumerate security products installed on the system from WMI", \\
  "Execution: Anomalous execution of VBScript file", \\
  "Execution: Attempt to download / exec with javascript / vbscript code", \\
  "Family: EICAR test sample", \\
  "Packer: Potentially unwanted application/program (VMProtect)", \\
  "Steal: Targeting Firefox Browser Password", \\
  "Steal: Targeting Internet Explorer Browser Password", \\
  "Steal: Targeting Opera Browser Password", \\
  "Steal: Targeting Outlook Mail Password", \\
  "Steal: Targeting Windows Saved Credential", \\
  "Stealth: Creating executables masquerading as browser clients", \\
  "Stealth: Creating executables masquerading as files from a Java installation", \\
  "Stealth: Creating hidden executable files", \\
  "Stealth: Modifying attributes to hide files"
 }}

 \section{Massive malware attacks in our dataset}
  \label{sec:massive}
  The following is the list of most frequently captured malware samples in our dataset.

\begin{table}[h]
\centering
\resizebox{\columnwidth}{!}{%
    \begin{tabular}{|l|c|c|c|c|c|}
    \hline
    \textbf{Attack} &  \textbf{Type} &  \textbf{A-count} & \textbf{B-count}  &  \textbf{Infection}     &  \textbf{What it does}   \\ \hline \hline
      
      Donoff         & Downloader & 28      & 445  &    Email    &   downloads Cerber    \\
      
      Skeeyah         & Trojan    & 55 & 207 &   Email     &   opens a backdoor    \\
      Tiggre         & Trojan  & 13  & 120     &   Malicious website     &   mines cryptocurrency    \\
    Madeba &  Trojan &   1 &    89  &     Email   & downloads another malware       \\
    Dynamer         & Trojan & 56   & 103     &    Dropped by other malware    &   downloads another malware     \\
    Fareit         & Spyware & 39   & 87     &      Dropped by other malware  &   steals sensitive information     \\
    Bluteal         & Trojan & 0   & 45     &   Dropped by other malware     &   gives remote access     \\
    Occamy         & Trojan & 2   & 37     &   Dropped by other malware     &   steals sensitive information     \\
    Locky         & Ransomware & 0   &    31  &   Email      & encrypts files      \\
    Nemucod         & Ransomware & 38   &    45  &   Email      & encrypts files      \\
 
\hline
    \end{tabular}
    }
    \vspace{5pt}
    \caption {Massive malware attacks in our dataset. \label{table:massive}}
\end{table}

\end{document}